\documentclass[10pt,conference] {IEEEtran}
\IEEEoverridecommandlockouts
\usepackage{cite}
\usepackage{amsmath,amssymb,amsfonts}
\usepackage{algorithmic}
\usepackage{graphicx}
\usepackage{textcomp}
\usepackage{xcolor}
\usepackage{colortbl}
\usepackage{breakurl}

\usepackage{multirow}
\usepackage{url}
\usepackage{subfig}
\usepackage{listings}
\usepackage{fancyvrb}
\usepackage{tcolorbox}
\usepackage{booktabs}
\def\BibTeX{{\rm B\kern-.05em{\sc i\kern-.025em b}\kern-.08em
    T\kern-.1667em\lower.7ex\hbox{E}\kern-.125emX}}
\begin{document}

\title{An insight into the technical debt-fix trade off in software
backporting\\
\thanks{Natural Sciences
and Engineering Research Council of Canada (NSERC)}
}

\author{
\IEEEauthorblockN{Jarin Tasnim, Debasish Chakroborti, Chanchal K. Roy, Kevin A. Schneider}
\IEEEauthorblockA{\textit{Department of Computer Science} \\
\textit{University of Saskatchewan}, Saskatoon, Canada \\
\{jarin.tasnim20, debasish.chakroborti, chanchal.roy, kevin.schneider\}@usask.ca}
}

\maketitle

\begin{abstract}
 Maintaining software is an ongoing process that stretches beyond the initial release. Stable software versions continuously evolve to fix bugs, add improvements, address security issues, and ensure compatibility. This ongoing support involves \textit{Backporting}, which means taking a fix or update from a newer version and applying it to an older version of the same software. As software versions evolve, new technical debt can arise during backport maintenance activities.
This study examines the technical debt involved in fixing 105,396 commits from 31,076 backport sources across 87 repositories in three software ecosystems (Apache, Eclipse, and Python). The goal is to identify when and why new technical debt arises during backporting in stable source code. Our results indicate that approximately 4.3\% of backports introduce new technical debt. Apache contributes the most absolute instances, while Python and Eclipse exhibit nearly three times higher debt-to-commit ratios than Apache. Feature migrations make older Apache releases debt-prone in the early phase, whereas Python and Eclipse releases tend to accumulate technical debt mostly during the middle phase of their release cycles. Additionally, developers who are inexperienced, under high workloads, or non-owners are more likely to introduce technical debt during backporting.


\end{abstract}

\begin{IEEEkeywords}
Backporting, Code Smells, Technical Debt, Long-term support, Software Maintenance.
\end{IEEEkeywords}

\section{Introduction}
Many software enterprises adopt source code versioning and provide critical maintenance support in parallel for older stable releases along with the upstream release \cite{b_m1}. In such cases, while the most upstream release delves toward future milestones, the old stable releases receive either long-term support (LTS) or short-term support (STS) \cite{url_lts,b_m1} during the release lifetime. In the industrial context, both the LTS and STS supports are reinforced by a standardized process called Backporting.   
It refers to the action of taking a patch developed for the newer release and porting it to an older release of the same software. Despite the additional costs and resources it demands, backporting remains a common practice in product-driven software \cite{url_lts, b2,b_m1}, as its benefits are twofold, addressing both stakeholder and user perspectives. From the stakeholder standpoint, this scheme helps preserve the foundations of information security \cite{b2, b_m2}. By addressing existing bugs and glitches over time, developers reinforce the targeted stable release against potential vulnerabilities throughout the support pipeline. At the same time, from a human-centric perspective, backporting enhances flexibility and dependability for users, many of whom refrain from upgrading to the latest release due to the inherent discomfort (time and effort constraints) for such upgrades \cite{l1,l2}.


Take the example of Windows 7 \cite{url_window}, released in 2009, which still holds a 22.56\% share of the desktop operating system market as of 2023.  Not just major operating systems like Windows and Linux distributions, but programming languages (Java, Python, Node, etc) \cite{url_lts,b_m2}, management systems (Collabora Suite, MS Suite, etc.), networking software (Cisco, Junitor, etc.), application frameworks (Django, Spring, Red Hat etc.)\cite{b_m1} and many other software systems require support for older releases. Old stable releases are actively maintained and subsidized by backports. 

\begin{figure}[bp]
\centering
\includegraphics[width=7.8cm]{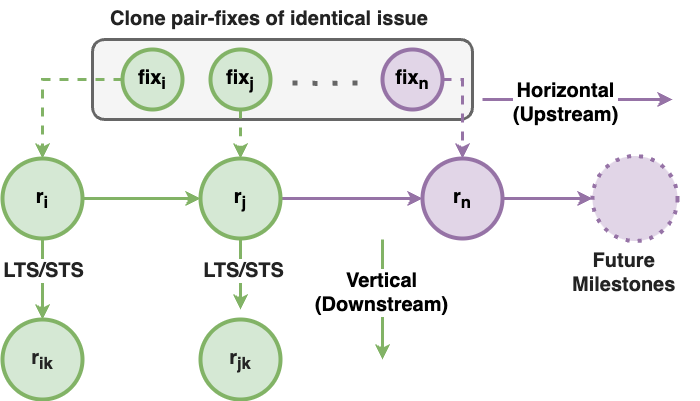}
\caption{Backport code propagation among different versions}
\label{fig:LTS}
\end{figure}

The systematic application of the backporting approach are performed with Version Control Systems (VCS) \cite{flynt2013characterizing, tasnim2023does, ss2} features. In widely used VCS such as Git, it is possible to directly port code from a newer release to an older one. However, this can include unnecessary changes when comparing two branches for a pull request \cite{b2}. To address this, a duplicate of the original pull
request is dispatched in the reverse tack that contains the patch
or its modified patch to make the fix compatible with old stable
releases.

To elaborate, taking a glimpse at Fig. \ref{fig:LTS}, the stable release versions $r_{i}$ and $r_{j}$ are actively maintained with the upstream repository $r_{n}$. When a new bug or critical issue appears, it is originally fixed with $fix_{n}$  in the most upstream (recent) imprint  $r_{n}$. This pull is referred to as \textit{Forwardporting} due to its alignment with the parallel code flow in the upstream direction. However, if the identical issue persists in older releases $r_{i}$ and $r_{j}$, the developers dispatch  duplicate pull requests (suppose $fix_{i}$ and $fix_{j}$) in the backward direction of the upstream repository to address the identical issue in the stable release $r_{i}$ and $r_{j}$, respectively. Thus, backports $fix_{i}$ and $fix_{j}$  are duplicate clones of original pull requests $fix_{n}$ with additional changes to ensure that fixes are compatible with the targeted release.

Developers often describe backport changes for older releases as ``Quick fix" or ``Hotfix" \cite{b_m0}, raising concerns about their safety and potential contribution to technical debt in the existing source code. Stable releases, as they progress through LTS and STS support cycles, are naturally susceptible to maintenance degradation. Backporting is motivated and executed differently from upstream development \cite{tasnim2023does, b_m1}, often without consistently following standardized inspection or review protocols. In addition, empirical evidence demonstrates that around 50\% of the backports need to incorporate on average 13\% of mutations beyond the original pull to make the patches compatible with the stable release structure \cite{b2, tasnim2023does}. Nevertheless, it remains uncertain when and why these mutation changes in backport are prone to introduce new technical debts to the stable releases.

Technical Debt (TD) is the additional cost and effort incurred in the future due to an expedient or suboptimal design\cite{td_0_Cunningham}.
Researchers have shown that TD \cite{yamashita2012code, td_2,s7} can lead to software bugs, slow performance, increased software failures, and other issues that degrade the user experience. As stable releases mature over time, the manifestation of new TD issues can directly impact existing user experience and trust. In addition, technical debt in backports opposes the fundamental objective of backports. Backports are typically considered a remediation or refactoring process for existing issues in the release. Hence, it is counterproductive to add additional issues while dealing with existing issues across releases. To our knowledge, no attempt has been made to investigate technical debt to fix trade-off in the backport maintenance.



Our study makes significant contributions to the understanding of technical debt in backporting practice through a comprehensive empirical analysis. 
The major contributions of this paper include the following:

(1) \textit{Empirical Analysis of Technical Debt:} We conducted a large-scale empirical study of technical debt in backporting, analyzing 105,396 backported commits across 87 open-source projects from the Apache, Eclipse, and Python ecosystems. Using SonarQube, a state-of-the-art TD analyzer, we found that approximately 4\% of backport mutation changes (4,549 identified instances) introduced technical debt\cite{replication_CASCON}.



(2) \textit{Insights into Release Evolution:} Our heuristics-based methodology provides quantitative evidence on how software ecosystems leverage backporting differently according to their specific needs, which in turn affects the timeline and accumulation of technical debt.

(3) \textit{Contradiction of Common Knowledge:} The common assumption is that backported changes are inherently safe \cite{b_m1, url_lts}, as they originate from tested upstream code and typically adhere to \textit{Feature Freeze}, a policy that restricts adding new functionality to release code. However, our findings indicate that this is not always the case. In ecosystems like Apache, which employ modular or plugin-based architectures, feature freeze is not always strictly enforced, which often leads to new technical debt.



\section{Related work}

In this section, we review relevant research that aligns with our objectives and explain how our research objective directs a unique perspective on software maintenance from two prime research domains: (1) backporting maintenance and (2) technical debt domain.

\subsection{Backporting Maintenance:}
Several studies have analyzed backporting practices, with a primary focus on the Linux operating system. These studies include automating backport recommendation \cite{tian2012identifying, thung2016recommending}, defining strategies \cite{b2, bogart2021and,ss2}, and detecting and characterizing porting errors more  efficiently\cite{ray2013detecting}. Tian et al. \cite{tian2012identifying} proposed automatically identifying bug-fixing patches in the Linux kernel to apply them to older long-term releases. Ray et al. \cite{ray2013detecting} characterize porting error traits to help developers deal with backports. Thung et al. \cite{thung2016recommending} suggested an automatic recommendation approach to choose potential code modifications that should be backported in Linux device drivers. FIXMORPH \cite{shariffdeen2021automated} is proposed by Shariffdeen et al. to automatically backport change patches from the upstream Linux kernel version into earlier stable versions. Using the syntactic structure to locate patch locations and applying transformation rules based on syntactic similarity to recommend backports. Shi et al. constructed a prototype for semantic patch backporting called SKYPORT \cite{shi2022backporting} and tested it on web application injection vulnerability patches. To understand and characterize the semantics of injection vulnerability patches, it established a logic representation for injection vulnerabilities. Yang et. al. \cite{yang2023enhancing} propose a patch type-sensitive approach to automatically backport OSS security patches, which rely on both patch type and semantics. Specifically, their backport automation tool TSBPORT outperforms state-of-the-art approaches, correctly recommending backports with 87.59\% of patches. In addition, a large-scale survey by Bogart et al. \cite{bogart2021and} involving over 2,000 developers across 18 ecosystems indicates that backporting is underutilized, as most developers report rarely investing additional time in backporting changes to their packages. Chakroborti et al. \cite{b2} \cite{10.1145/3639478.3643079} analyzed 68,424 backports from 10 GitHub projects and found that backporting is not limited to bug and security fixes in stable releases but also includes a wide range of test, documentation, and feature changes. Decan et al. \cite{ss2} examined backporting practices in major dependency package distributions (Cargo, npm, Packagist, and RubyGems) and found that while most security vulnerabilities affect multiple release series, they are typically only patched in the latest one.

\subsection{Technical Debt :}
The conceptual notion of technical debt (TD) was initially borrowed by Cunningham in 1992  \cite{td_0_Cunningham} from the financial concept of deficit to illustrate the accumulation of debt \cite{kruchten2012technical} that occurs when the emphasis is placed on saving time rather than ensuring optimal quality \cite{td_5, s7}. Simply put, technical debt refers to shortcomings in internal software quality that make future modifications and further development more challenging \cite{td_3}. Over the years, numerous studies have delved into how technical debts and quality concerns are introduced by developers as well as the duration for which technical debts like code smells \cite{td_3} and bugs tend to persist within a software system. In addition, the research community also looked into the probable side effects of technical debt existence \cite{td_side_effects} in source code and concluded TD concerns are responsible for an increase in change- and fault-proneness as well as the decline of software comprehension \cite{td_comprehension} and maintainability \cite{yamashita2012code}.
Khomh et al.\cite{s7} found that the presence of code smells and technical debt is associated with significantly higher change and fault proneness.  Tufano et. al. \cite{td_3} report that most of the time, code artifacts are affected by code smell instances since their creation.

Although the literature encompasses both cross-sectional \cite{alfayez2018empirical, klinger2011enterprise} and longitudinal approaches \cite{arcoverde2011understanding,  molnar2020longitudinal, nayebi2019longitudinal} to technical debt occurrences, they specifically focus on the upstream repository. In this study, we focus instead on the previous stable releases and their evolution during backporting. 
Molnar et al. \cite{td_7} investigated the long-term evolution of technical debts by analyzing the static source code of three projects across multiple releases, providing insight into how debt instances develop from the initial to the latest release. While this approach offers a horizontal perspective of software evolution across releases (Axes in Fig. \ref{fig:LTS}), it lacks the vertical perspective of evaluating individual release lifecycles. Although the negative impacts of TD on software performance and lifecycle are well known \cite{td_side_effects, td_comprehension}, little is understood about how, when, and why it emerges in LTS and STS release evolution. To address this, we identify and characterize 4,549 instances of technical debt within release evolution.



                                            


\section{Study Design}

\subsection{Research Questions}
The core intent of the project is to identify when and why technical debts manifest in backporting maintenance. To be precise, we aim to answer the following three research questions to dig deeper into the technical debt-fix trade-off in the backporting procedure.
\begin{itemize}
     
    \item \textbf{ RQ1: Does the process of backporting introduce new technical debt into stable releases?}

As stable releases age with time, the system experiences a decline in performance and a rise in failure occurrence rate where backporting convey the industrial effort to assure modest dependability and minimize threats in current business-driven and safety-critical software \cite{tasnim2023does, b_m2}. Thus, the changes incorporated in the LTS and STS are exclusively maintenance efforts to deal with the existing debt in release and hold the existing stable releases alive for users. As releases evolve, this lets us question whether or not they introduce new debt. To this purpose, we investigated the presence of technical debts in the backport history to release artifacts. 



\item \textbf{RQ2: When are technical debts introduced in release timeframes through backporting? }

Prior to a release launch, development teams often engage in rapid development to meet deadlines \cite{td_11, td_3}, which could inherently result in bugs \cite{release_time_pressure}, vulnerabilities, and quality concerns \cite{td_3}. As the release matures through bug and security fixes through backporting \cite{release_bug_1}, it progresses toward its \textit{End of Life (EOL)} when it is no longer supported or maintained. The anecdotal view is that it is less likely to accumulate any new technical debt through backport \cite{b_m1}. This research query aims to examine to what extent this conventional assumption is valid in release evaluation through backporting. Especially, we scrutinize \textit{when} technical debt instances are introduced in stable releases, is it probable to transpire at the initial stage of release maintenance or when the release has almost reached its maturity?


  \item \textbf{RQ3: Why do developers introduce new technical debt through backporting?}

Previous research has identified several attributes—such as deadline pressure \cite{td_3, td_10}, excessive workload, and lack of developer experience—as key factors contributing to the introduction of technical debt during artifact evolution. Thus, we aim to empirically investigate the impact of such potential attributes that might contribute to technical debt manifestation through backporting. In particular, we take into account the developer profile \cite{td_experience}, release and artifact characteristics \cite{s7, td_3} referred to by the existing literature to identify their potential influences. 
\end{itemize}

\subsection{Dataset and context selection}

The current academic position suggests that TD characteristics are closely correlated with the chosen software domain \cite{td_2}. To ensure data triangulation and capture a broad spectrum of application, we initiated our analysis with 100 random repositories from each of the two Java ecosystems (Apache\footnote{\url{https://github.com/apache/}} and Eclipse\footnote{\url{https://github.com/eclipse/}}) and 80 repositories from the Python\footnote{\url{https://github.com/python/}} ecosystem. Our decision to scrutinize these ecosystems is deliberate, aiming to encompass projects that exhibit particular characteristics, such as (i) diversity in project scope and sizes, e.g., Python repositories are generally smaller in scope than the Java ecosystems projects \cite{context_0}, (ii) distinction is code architectures \cite{context_1}, e.g., Eclipse ecosystem supports plug-in architectures, Apache consists libraries, frameworks, build tools, and Python supports third-party data analysis libraries for scientific computing (iii) distinction in social coding surroundings, e.g., Python projects are often developed by small teams \cite{context_0} whereas dozens of developers carry out several Apache projects \cite{context_2}, and (iv) Application language domain usage \cite{context_0}, e.g., Apache and Eclipse are in Java whereas we pick Python ecosystem to see whether debt instances reflect the distinct pattern in different languages source code.


To distinguish backports, we aim to identify how they are referred to within social coding environments (e.g., Git). Chakroborti et al. \cite{b2} \cite{chakroborti2024reback} observed that developers commonly include the keyword ``backport'' in the pull request metadata, including titles, tags, and descriptions. Accordingly, we performed a case-insensitive search for the keyword ``backport'' across all pull requests in 280 projects using the automated GitHub Web Search API, identifying 112 projects that contained the keyword. One of the authors then conducted a manual review of the git repositories and release notes for each of the 112 projects. We found that some projects used backporting solely to update dependencies by integrating third-party GitHub extensions (e.g., Dependabot). For this analysis, we intentionally excluded 17 such projects, as our focus is on technical debt arising from human-incorporated changes. Additionally, we excluded 8 projects whose release metadata and branch specifications were not adequately documented. Table \ref{tab:ecosystem_traits} summarizes the remaining 87 projects across the three ecosystems studied, detailing their size ranges (KLOC), the number of issues, the total number of backported commits, and the number of backports. All the 87 subject system use either JIRA\footnote{\url{https://www.atlassian.com/software/jira}} or BUGZILLA\footnote{\url{https://www.bugzilla.org/}} for issue tracking.



                                            


\setlength{\textfloatsep}{5pt plus 1pt minus 2pt}

\begin{table}
\centering
\resizebox{\columnwidth}{!}{%
\begin{tabular}{ccccccc}

\hline
 Ecosystem & \#Project & \#issues & \# commits & \#backport (PR) \\
 \hline
 Apache & 58    & 22341  & 82192 &25344 \\
Eclipse   & 12    & 12771  & 12765 & 3278 \\
Python & 17    & 10743  & 10439 & 2454 \\
\hline
Total & 87  & 45855 & 105396 & 31076 \\
\hline
                                            
\end{tabular}}
\caption{Data distribution of selected projects across ecosystems used for analysis}
\label{tab:ecosystem_traits}

\end{table}%







We acknowledge that relying solely on human-allocated metadata may introduce a considerable number of false positives in the dataset. To ensure that our dataset represents only true backports, we applied two constraints: (i) the pull request base (i.e., the destination branch where the changes are merged) must match a release branch of the repository, and (ii) the pull request must have been created within the corresponding release support timeframe. One of the authors manually identified and validated these release branches and their corresponding EOL date by inspecting git repositories and release notes during data collection. The replication package for this study is publicly available \cite{replication_CASCON}.



\subsubsection{\textbf{Debt-inducing commit identification by SonarQube}}


For the releases  $r\subset \{ r_{0}, r_{1},r_{2}... r_{n} \} $  of each project $P_{i}$, we scan the entire backport history each release $r_{i}$ receives with the SonarQube\footnote[3]{https://www.sonarsource.com/products/sonarqube/} to identify the commits that are responsible for TD manifestation. The choice of SonarQube for this study is motivated by two key factors: (i) it is a state-of-the-art TD analyzer for identifying code smell, bugs, and vulnerabilities with corresponding severity measure (ii) It offers flexibility by equipping the ``Plug and play" attribute to customize the scanner based on language, infrastructure and build configuration (e.g., Java projects can be on maven and build. As the context of data retrieved requires convergence on diversity, this feature directly aligns to achieve coverage.






Configuring \textit{SonarQube} required us to group the selected subject system based on the build manuscript and instantiate corresponding \textit{Sonar Scanner} for each group to communicate with the server. This involved configuring the \textit{Sonar Server} with PR decorator to check the commit patches and return responses. We configured the PR Decorator to report only TD issues that were not present in the pre-commit state; however, they appear after scanning the post-commit integration.

For each TD reported by \textit{SonarQube}, we consider the commit introducing the change as a \textit{Debt-inducing commit}. Thus, the debt-inducing commit sequence for each release $r_{i}$ is defined as $c \subset { c_{1}, c_{2}, c_{3}, \dots, c_{n} }$. To enhance reliability, the two authors validate the SonarQube outputs through qualitative checks and by discussing cross-verified results from other tool.



\begin{table}[htbp]
\centering
\resizebox{\columnwidth}{!}{
\begin{tabular}{cp{5cm}c}
\hline
 Tag & Description & Value \\
 \hline \rowcolor{lightgray}  \multicolumn{3}{c}{ Commit Goal} \\ \hline
Bug-fix   &   The commit fixes a bug in target release &  [Yes, No]\\
Security-fix & The commit fixes a security concern in release  &  [Yes, No] \\
Enhancement & The commit attempts to enhance the release code   &   [Yes, No]\\
Migration & The commit attempts to migrate feature into the release code   &   [Yes, No] \\
\hline \rowcolor{lightgray}  \multicolumn{3}{c}{Developer Attribute} \\ \hline
Workload & The work pressure on the developer    &  [Low, Mid, High] \\
Ownership & The commit author is the owner of the modified code artifacts & [Yes, No] \\
Exposure & The experience of the commit author working with the release    &  [Novice, Moderate, High] \\
\hline
\end{tabular}
}
\caption{Heuristics used for annotating technical debt–inducing commits}
\label{tab:table_commit_tag}
\end{table}

\subsubsection{\textbf{Annotating debt-inducing commits based on heuristics}}

Once debt-inducing commits have been identified, the next step is to understand when and why a technical debts are manifested in the release timeline. Employing the annotating instrument to investigate the underlying causation and pattern of artifact change has been a common practice\cite{chidamber1994metrics}. Particularly, we are inspired by the annotating directions of Tufano et. al. \cite{td_3}. Nevertheless, our contribution is novel compared to prior studies, as it applies these methods in the context of backporting. In addition, we fine-tune the mining characterization of tags to capture insights for release evolution. Among all the annotating mechanisms, the Maturity Phase (MP) tag is particularly designed to answer \textbf{When} (in $RQ_{2}$), while the rest of the tags (listed in Table \ref{tab:table_commit_tag}) are derived for responding to \textbf{Why} (in $RQ_{3}$).

\textbf{Maturity Phase (MP):}
To answer \textit{when} a debt-inducing commit is introduced in a release maintenance cycle, we aim to determine the phase of the release during which the TD issue occurred. Specifically, we investigate whether it is more likely to arise in the initial stage of release maintenance, mid-way through, or as the release approaches maturity. The formation and design of the \textit{Maturity Phase} heuristic were driven by the need to capture this understanding.

We utilized Git commands to capture the active period of maintenance support for stable releases from its corresponding git release branch. For debt-inducing commits merged to a stable release branch $r_i$, we use the git command: \texttt{git rev-list --merges --boundary \linebreak
--format="\%cd" r\_i}

\begin{figure}[htbp]  
\centerline{\includegraphics[width=8cm]{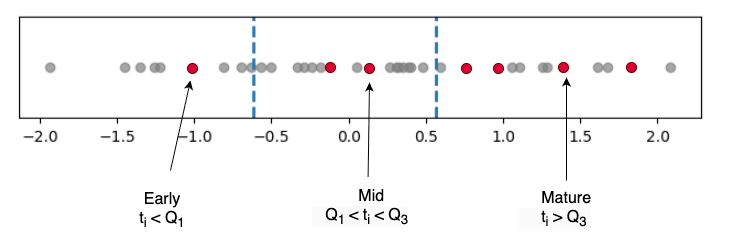}}
\caption{The design of Maturity Phase (MP) heuristics.}
\label{fig:study-2_rq2_maturity_phase}
\end{figure}

This command lists all merged commits in reverse chronological order from the forking point of the release branch $r_i$. We then recorded the commit dates of the first and last merged patches as the beginning and ending points (EOL) of maintenance support, denoted as $r_{start}$ and $r_{stop}$, respectively. In addition, we manually inspected the validity of our measure as the active maintenance life span of the release branch by cross-checking its consistency with release documents. We then try to locate when debt-inducing commits manifest in the quartile distribution of release maintenance lifespan.
Given $Q_{1}$ and $Q_{3}$ being the first and third quartiles of release support window of $\{ r_{start} , r_{stop} \}$, we compare the debt-inducing commit time $t_{i}$, and assign the Maturity Phase heuristic as \textit{Early} if $t_{i} \leq Q_{1}$, \textit{Mid} if $Q_{1} \leq t_{i} \leq Q_{3} $ and \textit{Mature} if $ t_{i} \geq Q_{3}$. Figure \ref{fig:study-2_rq2_maturity_phase} presents the MP criterion of debt-inducing commits (marked with red color) in a release life span of Apache Storm project. 

\textbf{Commit Goal}: To understand the purpose of each change, we employ \textit{Commit goal} heuristics, which help characterize what the developer was performing (e.g., whether a commit was a bug fix or a feature enhancement) when new technical debt was introduced.



To identify \textit{commit goal}, we implement the regular expression-based technique described by Fischer et al. \cite{td_6} to trace potential links between commits to issues. In this approach, we first extracted the $issue\_id$ by mining regular expressions in the backport body or the commit note, (e.g., ``fixed issue \#ID" or ``issue ID"). Secondly, for each $issue\_id$ corresponding of a debt-inducing commit, we extracted the corresponding issue reports from the REST API of the issue-tracking system. We then use the $issue\_type$ (i.e., whether the backport issue resolves a bug, security, enhancement, or feature) from the retrieved data and classify the debt-inducing commits into the four classes of backport fixes shown in Table \ref{tab:table_commit_tag}. The first author manually inspects each debt-inducing commit, comparing it with the version-control log, and confirming the validity of this tag for the entire dataset. 


\textbf{Developer Attributes: }
Developer profile \cite{td_11}, experience \cite{td_experience}, and work pressure \cite{td_3} have been identified as key factors influencing the accumulation of technical debt in source code. Accordingly, we investigate how a developer’s workload, code ownership, and exposure to specific code artifacts within a release affect the propagation of technical debt during the backporting process.

The \textit{Workload} heuristics intend to capture the level of work pressure the developer had when they induced the technical debt in backport. This heuristic regards the contribution of an individual in a time window of one month with respect to the number of LOC modified by the contributor in the release branch $r_{i}$ (starting from the the day initial commit is recorded). When a developer introduces a new debt in a month (denoted as $m$), we estimate the workoad distribution for all developers in the repository at that time and assign values \textit{Low} as workload  if the developer (d) inducing the  debt  had a $workload \leq Q_{1}$, \textit{Medium} if $Q_{1} \leq workload \leq Q_{3} $ and \textit{High}  if $ workload \geq Q_{3}$ where $Q_{1}$ and  $Q_{3}$ being the first and the third quartile on distribution scale . We acknowledge that this measure is an approximation because distinct commits may require contrasting amounts of LOC change, which might not reflect the actual cognitive complexity of the task. In addition, an open-source contributor may simultaneously work on other projects. However, this parameter helps characterize debt-inducing commits, ranging from extensive to minimal contributions, and provides an abstract overview of their influence on debt embodiment \cite{td_3}.

\begin{table*}
\centering
\resizebox{\textwidth}{!}{
\begin{tabular}{@{}ll ccccc c cc c @{}}
\toprule
  \multirow{2}{*}{ Ecosystem } &  \multirow{2}{*}{Type} & \multicolumn{5}{c}{Severity} &  Overall & TD/Commit &  Top rule violation  \\
\cmidrule(lr){3-7} 
  & &  \shortstack{Blocker\\\includegraphics[width=0.5cm]{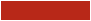}}
  &  \shortstack{ Critical \\\includegraphics[width=0.5cm]{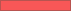}} 
  & \shortstack{Major \\\includegraphics[width=0.5cm,height=0.1cm]{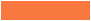}}
  & \shortstack{Minor \\\includegraphics[width=0.5cm,height=0.1cm]{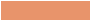}}
  & \shortstack{ Info \\\includegraphics[width=0.5cm]{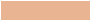}} & (count)&  (\%) \\
\midrule
      \multirow{3}{*}{Apache}  & Bug  & \multicolumn{5}{l}{\includegraphics[width=1.5cm,height=0.18cm]{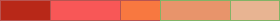}} & \multirow{3}{*}{2441} & \multirow{3}{*}{2.9} &   cwe, overflow, cert\\
                              & Vulnerability & \multicolumn{5}{l}{\includegraphics[width=1.1cm,height=0.2cm]{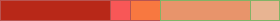}} &&   &
                              cert, cwe\\
                              & Code smell  & \multicolumn{5}{l}{\includegraphics[width=6.5cm,height=0.2cm]{Image/rq1/1.png}} & & & design, pitfall, brain-overload\\
    \hline                  
      \multirow{3}{*}{Eclipse} & Bug  & \multicolumn{5}{l}{\includegraphics[width=0.6cm,height=0.18cm]{Image/rq1/2.png}} &\multirow{3}{*}{974}  & \multirow{3}{*}{9.0} &  symbolic-execution,  spring \\
                              & Vulnerability & \multicolumn{5}{l}{\includegraphics[width=1.8cm,height=0.18cm]{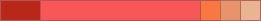}} && & injection, cwe, cert\\
                              & Code smell  & \multicolumn{5}{l}{\includegraphics[width=2.8cm,height=0.18cm]{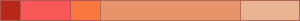}} & & &unused\\
                             
     \hline
      \multirow{3}{*}{Python} & Bug  & \multicolumn{5}{l}{\includegraphics[width=1.1cm,height=0.18cm]{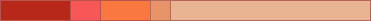}} & \multirow{3}{*}{1134} & \multirow{3}{*}{9.1} & quick-fix, cwe \\
                              & Vulnerability & \multicolumn{5}{l}{\includegraphics[width=0.8cm,height=0.18cm]{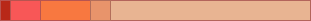}} & & &
                              cwe, injection, privacy\\
                              & Code smell  & \multicolumn{5}{l}{\includegraphics[width=3.4cm,height=0.18cm]{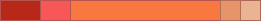}} & & & cwe, regex, performance\\
\bottomrule
\end{tabular}
}
\caption{Distribution of technical debt–inducing commits by type and severity across ecosystems}
\label{tab:sonarQube_report}
\end{table*}
The \textit{Ownership} tag is assigned to identify whether the developer introducing the debt is the owner of the files where the debt instance was detected. We compute whether or not the debt-inducing commit author holds the \textit{Ownership} of the code components being modified by the commits by following the ownership model by Bird et. al. \cite{td_4}. If $N = \sum^{j}_{i=1}commits$ is the total number of commits to the code artifacts, where the developer $d_i$ contributed with $N_{d} = \sum^{j}_{i=1}commits(d_{i})$, then the ownership of developer,  $Ownership(d_{i})$ can be defined as follows:

\begin{equation}
Ownership(d_{i}) = 
        \begin{cases}
            Yes & \text{if $\frac{N_{d}}{N} \geq \frac{N - N_{d}}{N} $ } \\
            No & \textit{Otherwise} \\
        \end{cases}
\end{equation}

Lastly, the \textit{Exposure} tag is assigned to identify whether the developer who is introducing the debt has previous experience working with the release branch $r_{i}$. The motivation behind the tag is to capture the cognitive awareness of the developer with respect to the changes they are incorporating. Palomba et. al. concluded developer experience and domain knowledge in code artifacts are critical attributes that could drive the developer conscious of quality design issues \cite{tag_1}. Considering developer is accountable for LOC change $l_{i}$ in a release branch $r_{i}$, the tag is allocated \textit{High} if the $l_{i}>=0.5N$, \textit{Moderate} if $l_{i}<= 0.5N$ and \textit{Novice} if $l_{i} = 0$ where $N$ is the total LOC of the release.

\begin{equation}
Exposure(d_{i}) = 
        \begin{cases}
            High & \text{if $ l_{i}>=0.5N $ } \\
            Moderate & \text{if $l_{i}<= 0.5N $ } \\
            Novice & \text{$l_{i} = 0$} \\
        \end{cases}
\end{equation}

Once all the tags have been assigned following the conveyed heuristics to the 4549 debt-inducing commits, We convey descriptive statistics mapping the density of commits distributed to each tag category in order to assess the results (Presented in Table \ref{tab:table_commit_goal}, \ref{tab:table_release} and \ref{tab:table_developper} ).

\section{Results and Discussion}
In order find answers to our research questions, the analysis of the outcomes is reported in this part.

\textbf{Answering RQ1: Does the process of backporting introduce new technical debt into stable releases? }


The analysis of technical debt across three software ecosystems—Apache, Python, and Eclipse—reveals distinct patterns. Table \ref{tab:sonarQube_report} shows the average issue types and their severity levels. Apache, with the largest number of backport commits (82,192) and technical debts (2,441), exhibits the lowest TD per commit ratio (2.97\%), indicating that its commits introduce relatively less technical debt on average. In contrast, Python (12,765 commits, 1,154 TDs) and Eclipse (10,439 commits, 954 TDs) show much higher TD per commit ratios is 9.04\% and 9.14\%, respectively, suggesting a higher density of technical debt per commit. In general, across all ecosystems, 4,549 technical debts were found in 105,396 commits, resulting in a TD per commit ratio of 4.32\%.

According to our analysis, the majority of technical debt was attributed to the maintainability domain, particularly code smells. 
For all three ecosystems, we noticed that the code smell types are prevalent debts compared to bugs and vulnerabilities. This trend was consistent across the various target applications and their historical data. 


In addition, we can also see the top rule tags (in Table \ref{tab:sonarQube_report}) assigned to the debt instances, as they give an intuitive notion of the major rule breaks for each type of technical debt. The major rules violated in the Apache ecosystem were Common Weakness Enumeration (CWE), memory allocation and buffer issues (overflow),  Security Category (cert, design, pitfall, and brain overload). On the contrary, we observe that most of the technical debts in the Eclipse ecosystem originate due to the violation of symbolic execution, spring, and injection rules. For the Python ecosystem, the top violated rules were  CWE, injection, privacy, regex, and performance.
\begin{tcolorbox}[colback=black!1!white,colframe=black!1!black]
While larger ecosystems like Apache may accumulate more technical debt in absolute terms, smaller or less active ecosystems tend to have denser technical debt, posing greater potential challenges for maintenance per code change.
\end{tcolorbox}


\textbf{Answering RQ2: When do developers introduce technical debts through backporting? }\\
Table \ref{tab:table_release} shows the mapping of \textit{Maturity Phase} (MP) heuristics to debt-inducing commits. Our analysis shows that early and mid-release maintenance phases are more prone to technical debt-inducing commits than the mature phase. In total, 1,981 commits (44\%) occurred in the early phase, 1,688 (37\%) in the mid phase, and 880 (19\%) in the mature phase.

\begin{table}[htbp]
\centering
\begin{tabular}{@{} c ccc  c @{}}
\toprule
 \multirow{2}{*}{ Ecosystem} &   \multicolumn{3}{c}{Maturity Phase} &  \multirow{2}{*}{ P-value} \\
\cmidrule(lr){2-4}  
  & Early & Mid & Mature \\ 
\midrule
\addlinespace[2ex]
      Apache   & \shortstack[c]{1632\\66.8\%} & \shortstack{549\\22.5\%} & \shortstack{260\\10.6\%} & 0.733\\
                            
    \midrule                
      \addlinespace[2ex]
      Eclipse  &  \shortstack{112\\11.4\%} &  \shortstack{ 573 \\58.8\%}  &  \shortstack{289 \\29.6\%} & \cellcolor{blue!15}0.027 \\
     \midrule
     \addlinespace[2ex]
      Python &  \shortstack{237\\20.9\%} &  \shortstack{566\\49.9\%}  &  \shortstack{331\\29.2\%} & \cellcolor{blue!15} 0.034 \\
     \midrule
     Total & 1981 & 1688 & 880 & - \\
\bottomrule
\end{tabular}
\caption{Distribution of debt-inducing commits by  \textit{Maturity Phase} heuristics}
\label{tab:table_release}
\end{table}

Across the three ecosystems, Apache repositories are more prone to technical debt in the early-release phase, with approximately two-thirds (66.8\%) of debts occurring during this period. In contrast, Eclipse and Python releases accumulate more debt in the mid-release phase, accounting for 58.8\% and 49.9\% of technical debts, respectively. In the mature phase, both Eclipse and Python acquire around one-third of their total debt, while Apache accounts for only 10\%. These findings challenge the common assumption \cite{release_bug_0, td_0_Cunningham} that software releases become increasingly stable over time and are less likely to introduce new issues in the mature phase.

To investigate the relationship between release maturity phases and debt manifestation, we analyzed the correlation between the MP categorical variables of debt-inducing and non-debt commits. Since measuring MP heuristics for all non-debt commits is computationally expensive, we applied a random stratified sample of 10\% of non-debt commits from the three ecosystems as strata. This sample included 11,793 non-debt commits, satisfying a 95\% confidence level with a 5\% margin of error. Using a \textit{Chi-Squared test} to assess the independence of these categorical variables (Table \ref{tab:table_release}), we found statistically significant results ($p < .05$) for both Eclipse and Python, but no significant correlation for Apache.

\begin{tcolorbox}[colback=black!1!white,colframe=black!1!black]
Technical debts in Apache projects tend to occur early in the release lifespan, while for Eclipse and Python, they mostly arise during the mid-phase of release maintenance. Release maturity significantly affects debt manifestation in Eclipse and Python, but not in Apache.
\end{tcolorbox}

\textbf{Answering RQ3: Why do developers induce new technical debt by backporting? }\\


\textbf{Commit goal:} After classifying the debt-inducing commits reported by SonarQube into four major categories, we analyze the percentage of debt-inducing commits in each category. The Apache ecosystem is more likely to introduce new technical debts through feature migration. When a feature is developed for an upstream release\cite{b_m0}, the developers often perform cross-branch backporting for old stable releases in Apache ecosystem to make the same feature available for their users. Around 55.2\% of technical debts are introduced by such feature migration of cross-release backports.

\begin{table}[htbp]
\centering
\resizebox{\columnwidth}{!}{%
\begin{tabular}{@{} l cccc c@{}}
\toprule
   \multirow{2}{*}{ Ecosystem }  &\multicolumn{4}{c}{Commit Goal } &   \multirow{2}{*}{ P-value }\\
\cmidrule(lr){2-5} 
\addlinespace[1ex]
  & \shortstack{ Bug-fix\\(\%)} & 
  \shortstack{Security-fix\\(\%)}  & 
  \shortstack{Enhancement\\(\%)} &
  \shortstack{Migration\\(\%)}  \\
\midrule
  Apache   & 15 & 13 & 16.4 & 55.2 & \cellcolor{blue!15}0.0362\\

      \addlinespace
Eclipse & 13 & 14 & 41  & 32 & \cellcolor{blue!15} 0.045\\
                             
     \addlinespace
Python  & 37 & 35.7 & 22.8  & 4.5 & 0.623\\
\bottomrule
\end{tabular}
}
\caption{Percentage (\%) distribution of debt-inducing commits by \textit{Commit-goal} heuristics.}
\label{tab:table_commit_goal}
\end{table}

To investigate the underlying causes of debt-inducing backports, we manually inspected a subset of backports. We observed that most debt-inducing commits, particularly those associated with code smells, originated from feature migrations. For instance, Apache Kafka \# 12327 backport performed an extensive feature migration to the target release branch that involved changing 31 files. There are multiple instances, where a significant change is propagated by backporting. To sum up, the developer community in Apache, as a multi-domain architecture\cite{context_2}, does not simply restrict  backporting practices in security or bug fixes, but also extensively utilizes it for feature migration across distinct releases which are prone to new technical debts. This contrasts with industrial expectations \cite{tasnim2023does}, which view backporting primarily as a mechanism for debt remediation. In the Apache ecosystem, 15\%, 13\%, and 16.4\% of technical debt were introduced during bug fixing, security fixing, and enhancement activities, respectively.

In the Python ecosystem, we identified 1,134 instances of technical debt. Most of these originated from developers’ efforts to address security concerns (35.7\%) and bug fixes (37\%), followed by enhancement activities (22.8\%). A large share of debt-inducing commits was observed in CPython, the reference compiler of the Python language, written in C and Python. For example, backport \#110223, merged into the 3.9 release branch, fixed a parser crash in the file ``pegan.c". However, the quality benchmark in C language considers designing double pointer data structures more problematic to understand and use correctly. Thus, Sonarqube denotes an issue entitled ``Object declarations should contain no more than two levels of pointer indirection" with ``brain overload" and ``pitfall" tags.

A total of 974 technical debts manifested in Eclipse ecosystem. According to our analysis, the majority of the technical debts in this ecosystem are introduced for enhancement followed by migration commits, which account for 41\%  and 32\% respectively. The bug-fixing and security- fixing commits are responsible for the rest of the 13\% and 14\% of TD issues in this ecosystem.

We further analyzed the relationship between technical debt occurrence and commit goals, as summarized in Table \ref{tab:table_commit_goal}. The Chi-Squared test revealed a statistically significant association $(p < .05)$ in the Apache and Eclipse ecosystems, whereas no such association was observed in the Python ecosystem.

\begin{tcolorbox}[colback=black!1!white,colframe=black!1!black]
Approximately 50\% of the technical debt in Apache projects originates from cross-release feature migrations, where the feature freeze policy is not consistently maintained. The majority of technical debt-inducing commits in Eclipse projects are also related to feature enhancement and migration. On the other hand, bug and security fixing activities in Python projects are prone to technical debts.
\end{tcolorbox}

\textbf{Developer traits:} Table \ref{tab:table_developper} presents the distribution of debt-inducing commits across the three developer heuristics. The breakdown indicates that workload negatively impacts developer contributions and can contribute to debt manifestation during backporting. On average, across all three ecosystems, 77.5\% of backports performed by developers with a higher workload were more likely to introduce new debt. To assess the statistical significance of this relationship, we conducted Chi-Squared tests on the workload categories. The resulting P-values for Apache (0.0211), Eclipse (0.0091), and Python (0.000612) suggest that workload is a significant factor in all three ecosystems and is likely to influence technical debt manifestation.

\begin{table}[htbp]
\centering
\resizebox{\columnwidth}{!}{%
\begin{tabular}{@{} c ccc ccc ccc @{}}
\toprule
  Ecosystem  & \multicolumn{3}{c}{Workload} &  \multicolumn{2}{c}{Ownership} & \multicolumn{3}{c}{Exposure} \\
\cmidrule(lr){2-4} \cmidrule(lr){5-6} \cmidrule(lr){7-9}
  & L & M & H & Yes & No & H & M & N\\ 
\midrule
      Apache & \cellcolor{blue!15} 15.6 & \cellcolor{blue!15}10.2 & \cellcolor{blue!15} 74.9 &
      42.6 & 57.4 & 
      26.3 & 14.2  & 59.5\\
                            
      \addlinespace
      Eclipse &  \cellcolor{blue!15} 9.2 &  \cellcolor{blue!15} 6.4 &  \cellcolor{blue!15} 84.4  & 
      32.6 & 67.2 & 
      \cellcolor{blue!15} 4.3 &  \cellcolor{blue!15} 19.6 &  \cellcolor{blue!15}76.1\\
                             
     \addlinespace
      Python    & \cellcolor{blue!15} 7.8 & \cellcolor{blue!15} 16.9 & \cellcolor{blue!15} 75.3  & 
      38.7 & 61.3 &
      \cellcolor{blue!15} 18.6 &  \cellcolor{blue!15} 4.2 &  \cellcolor{blue!15}77.2\\
\bottomrule
\end{tabular}
}
\caption{The percentage (\%) distribution of debt-inducing commits across developer heuristics. (The colored cells denote a statistically significant (P-value $= 0 < 0.05$))\\
\textit{Workload} (L - Low, M - Medium, H - High), \textit{Ownership} (Y - Yes, N - No) \textit{Exposure} (Novice- N, Moderate-M, High-H). }
\label{tab:table_developper}
\end{table}

From Table \ref{tab:table_developper}, the proportion of debt-inducing commits by non-owners was higher than owners for all three ecosystems. Examining the distribution of ownership, the distribution shows that approximately two-thirds of the technical debts in backporting practice are introduced by non-owners. The Chi-Squared significance testing of ownership categorical variables does not reveal any significant difference  (P-value $= 0 > .05$) between the debt-inducing commits and non-debt commits.

The \textit{Exposure} heuristics characterize developer's experience working with the release source code. According to our analysis,  developers with no previous experience working with the release are more likely to introduce technical debts for all three ecosystems. We can observe that Novice developers are responsible for approximately 59.5\%, 76.1\%, and 77.2\% of technical debts in the Apache, Eclipse and Python ecosystem respectively. Although the statistical significance testing of Exposure categorical variables does not reveal any significant difference between the debt-inducing commits and non-debt commits for Apache ecosystem ($\alpha= 0.73$), we found a strong correlation for Eclipse  ($\alpha= 0.041$) and Python  ($\alpha= 0.00653$) ecosystem

\begin{tcolorbox}[colback=black!1!white,colframe=black!1!black]
Developers who are inexperienced, under high workloads, or non-owners are more likely to introduce technical debt during backporting.
\end{tcolorbox}

\section{Lesson Learned}
This analysis provides useful insights and lessons for both the research community and industry practitioners:

\textbf{Lessons learned 1:} Although conventional wisdom suggests that long-term release support primarily serves to maintain stable releases, our study shows that in 4.3\% cases, backports introduce new technical debt, including bugs and vulnerabilities, into the release source code. Thus, it emphasizes the significance of executing quality checks at code propagation through backporting in release source code. In addition, the technical debt detection tools should consider the strategic standard of the backporting process. So that developers can get an intuitive assertion during the backport commit transition, particularly to prevent or mitigate the manifestation of new debt while dealing with existing debt in LTS and STS. 

\textbf{Lessons learned 2:} Software ecosystems leverage backporting differently based on their specific needs, which in turn affects the timeline and concentration of technical debt. Apache and Eclipse use backports mostly for feature migration, whereas the Python development communities mainly use backports for stable release necessities. These behaviors vary across ecosystems.

\textbf{Lessons learned 3:} Non-owners and developers with high workloads are more likely to introduce technical debts in backporting. This underscores the importance of rigorous code inspection, particularly when developers work under high workload or lack prior experience with release code. Understanding these factors allows managers and team leads to make more informed decisions and enforce backporting policies effectively. Given the significant influence of workload and exposure on TD manifestation, backporting policies should account for these criteria, and managers should monitor contributions from developers with limited release experience to guide subsequent maintenance decisions.

\textbf{Lessons learned 4:} We found technical debts are more likely to appear in Apache projects early in the release life cycle. On the other hand, Python and Eclipse projects receive the majority of the technical debts in the middle phase of the release maintenance time frame. Furthermore, 19\% technical debts at the mature phase of the release life span. We recommend Managers and team to align quality governance and maintenance strategies with each ecosystem’s debt-prone phases.

\section{THREATS TO VALIDITY}

Despite our efforts to ensure rigor, the following threats to validity remain:

\textbf{Construct validity}: The study relies on SonarQube for detecting and quantifying technical debt. While SonarQube \cite{sonar_1} is widely adopted in both research and practice, its rule set and detection strategies are not exhaustive and may miss or misclassify certain issues. To mitigate this threat, we used the latest SonarQube LTS with an industry-recognized default rule set to ensure consistency, and we cross-validated flagged issues against project documentation when necessary. 

\textbf{Internal validity}: Several steps in our data collection and analysis required manual intervention, such as validating debt-inducing commits and resolving ambiguities in release documents. These manual steps introduce the possibility of human error and subjective judgment. To limit this threat, we applied a double-checking procedure in which the second author independently reviewed manual classifications and resolved disagreements with the first author through discussion, thus reducing individual bias. 

\textbf{External validity}: The generalisability of our findings is constrained by both the studied systems and the chosen methodology. The selected projects and release branches may not represent all ecosystems or development practices. To mitigate this, we selected multiple projects with diverse characteristics (e.g., domain, size, and community activity) and clearly documented our selection criteria, allowing others to replicate the study or extend it to additional contexts.  Moreover, static analyzers often disagree with each other, as their detection rules, precision levels, and assumptions vary\cite{sonarQube_2}. Thus, we acknowledge that using PMD or Checkstyle could yield different outcomes, and future work could explore multi-tool validation.

\textbf{Conclusion validity}: We acknowledge that the Maturity cycle and Workload heuristics are designed on approximations to capture general traits, which may not always reflect real-world scenarios. For example, using a quartile-based scheme may not fully capture release maturity, as release decisions and strategies are human-driven. Similarly, computing a developer’s activity based on the Workload heuristic may be biased if a developer contributes to multiple projects within or outside the ecosystem. However, measuring the Workload of a developer outside of the ecosystem scope will raise a new bias by inflating or underestimating the measure.



Despite these limitations and inherent human biases, we argue that these heuristics provide a reasonable abstraction of their intended concepts. 



\section{Conclusion and future work}

Our result suggests while Apache ecosystem accrue more technical debt in total, Python and Eclipse ecosystem projects being less active in backporting tend to have technical debt density, which can make maintenance more challenging per code change. The common understanding of maintaining feature freeze in backporting is often not valid for a plugin-based ecosystem like Apache, which may explain why Apache release artifacts accumulate the highest absolute number of technical debt instances. About half of technical debts in Apache projects are related to cross-release feature migration. The majority of technical debt-inducing commits in Eclipse projects are also related to feature enhancement and migration. On the other hand, bug and security fixing activities in Python projects are prone to technical debt. 

In addition, technical debts are more like to occur in the early release maintenance in Apache ecosystem. However, for Eclipse and Python projects the density of technical debt manifestation is higher in the mid-phase of release life cycle. Upon further examination of the developer attribute, we discovered that  technical debts in backporting process is more likely to be introduced by inexperienced developers and developers with heavy workloads. This emphasizes the necessity to improve community collaboration and code quality inspection when extensive changes are made through backporting  procedures. 

\section{ACKNOWLEDGMENT}

This research is supported in part by the Natural Sciences and
Engineering Research Council of Canada (NSERC) Discovery grants,
and by an NSERC Collaborative Research and Training Experience
(CREATE) grant, and by two Canada First Research Excellence
Fund (CFREF) grants coordinated by the Global Institute for Food
Security (GIFS) and the Global Institute for Water Security (GIWS).


\bibliography{sample-base}

\end{document}